\documentclass[reprint,aps,prl,superscriptaddress,noeprint]{revtex4-1}

\usepackage[normalem]{ulem}
\usepackage{amsmath}
\usepackage{amsbsy}
\usepackage{amssymb} 
\usepackage{graphicx}
\usepackage{float}
\usepackage[caption=false]{subfig}
\usepackage{hyperref}
\hypersetup{
	colorlinks=true,       
	linkcolor=blue,          
	citecolor=green,       
	filecolor=magenta,      
	urlcolor=blue          
}

\usepackage[dvipsnames]{xcolor}

\begin{document}
\title{Asymmetric blockade and multi-qubit gates via dipole-dipole interactions}	

\author{Jeremy T. Young}
\email[Corresponding author: ]{jeremy.young@colorado.edu}
\affiliation{Joint Quantum Institute, NIST/University of Maryland, College Park, Maryland 20742 USA}

\author{Przemyslaw Bienias}
\affiliation{Joint Quantum Institute, NIST/University of Maryland, College Park, Maryland 20742 USA}

\author{Ron Belyansky}
\affiliation{Joint Quantum Institute, NIST/University of Maryland, College Park, Maryland 20742 USA}
\affiliation{Joint Center for Quantum Information and Computer Science, NIST/University of Maryland, College Park, Maryland 20742 USA}

\author{Adam M. Kaufman}	

\affiliation{JILA, University of Colorado and National Institute of Standards and Technology, and Department of Physics, University of Colorado, Boulder, Colorado 80309, USA}

\author{Alexey V. Gorshkov}
\affiliation{Joint Quantum Institute, NIST/University of Maryland, College Park, Maryland 20742 USA}
\affiliation{Joint Center for Quantum Information and Computer Science, NIST/University of Maryland, College Park, Maryland 20742 USA}

\date{\today}

\begin{abstract}

Due to their strong and tunable interactions, Rydberg atoms can be used to realize fast two-qubit entangling gates. We propose a generalization of a generic two-qubit Rydberg-blockade gate to multi-qubit Rydberg-blockade gates which involve both many control qubits and many target qubits simultaneously. This is achieved by using strong microwave fields to dress nearby Rydberg states, leading to asymmetric blockade in which control-target interactions are much stronger than control-control and target-target interactions. The implementation of these multi-qubit gates can drastically simplify both quantum algorithms and state preparation. To illustrate this, we show that a 25-atom GHZ state can be created using only three gates with an error of 7.8\%.

\end{abstract}


\maketitle

\textit{Introduction}.---
Strong and tunable interactions between Rydberg states have positioned neutral atoms as a versatile platform for quantum information science and quantum simulations. Many of these proposed applications rely on Rydberg blockade, a process in which a single Rydberg excitation prevents nearby atoms from being excited to the Rydberg state. In recent years, there have been extensive efforts to characterize and improve the performance of entangling two-qubit gates based on Rydberg blockade, first proposed in Ref.~\cite{Jaksch2000} and further investigated in Refs.~\cite{Saffman2005,Saffman2010,Xia2013}. This novel approach was later followed by a variety of theoretical extensions \cite{Brion2007a,Goerz2014, Muller2011,Muller2011a,Muller2014,Moller2008,Rao2014,Beterov2016, Shi2014, Shi2017, Tian2015, Wade2016, Wu2010a,Petrosyan2014, Tian2015} and experimental implementations \cite{Isenhower2010,Wilk2010,Jau2016,Zeng2017,Picken2019}. Recently, two-qubit entangling gates have been realized experimentally with high fidelities \cite{Levine2019,Graham2019}.

Importantly, the long-range character of Rydberg van der Waals (vdW) and dipole-dipole interactions opens the possibility of engineering entangling gates involving many qubits. Although two-qubit entangling gates are sufficient for universal quantum computing, multi-qubit entangling gates can provide significant speedups for quantum algorithms and state preparation. For example, multi-target Rydberg gates~\cite{Muller2009,Su2018} enable the implementation of Shor's algorithm in constant time \cite{Hoyer2005}. Conversely, multi-control Rydberg gates \cite{Isenhower2011, Gulliksen2015, Su2018a} allow for efficient implementations of Grover's search algorithm \cite{Molmer2011}.

\begin{figure}
\includegraphics[scale=.55]{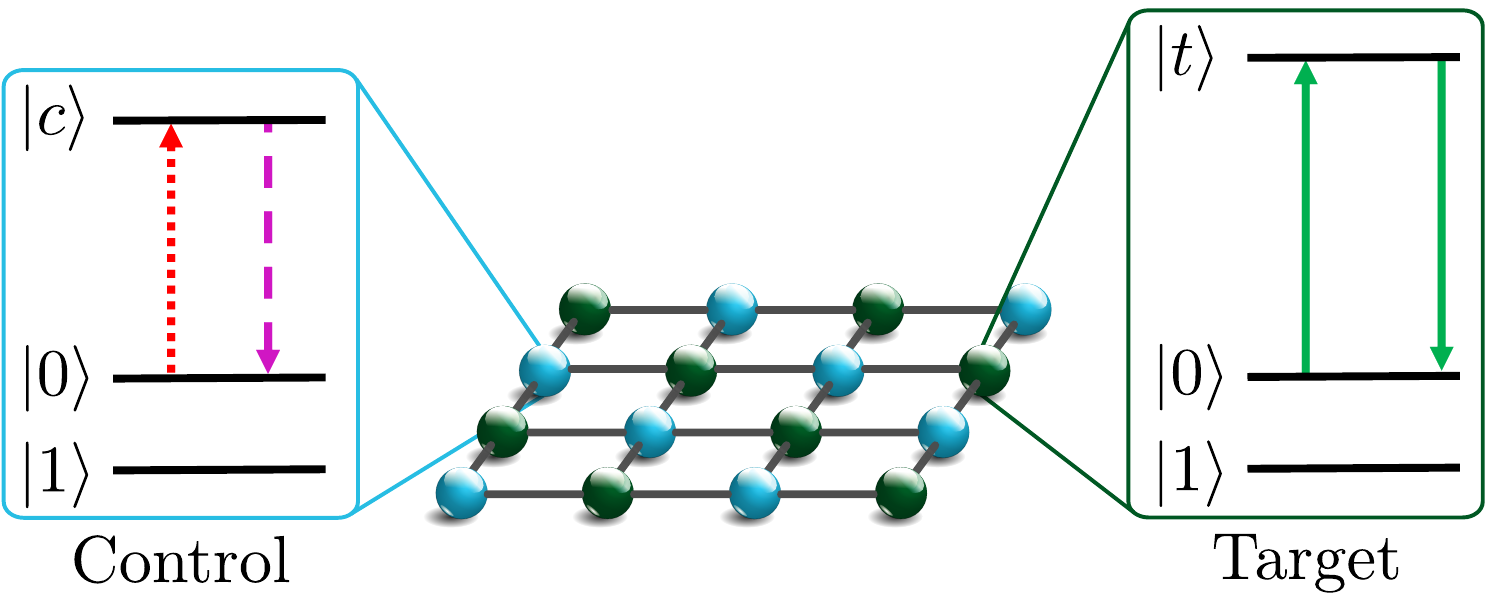}
\caption{Pulse sequence to realize controlled-Z gates, where light blue (dark green) spheres represent control (target) atoms. Other configurations of control and target atoms are possible. (1) A $\pi$ pulse excites the control qubits in the $|0\rangle$ state to the $|c\rangle$ state, indicated by the dotted red arrow. (2) A $2 \pi$ pulse through the $|t\rangle$ state is applied to the $|0\rangle$ state of the target qubits, which is blockaded if any control qubits are in the $|0 \rangle$ state, indicated by the solid green arrows. (3) A $-\pi$ pulse de-excites the control bits from the $|c\rangle$ state to the $|0\rangle$ state, indicated by the dashed purple arrow. \label{Pulses}}
\end{figure}

The conventional implementation of the two-qubit Rydberg-blockade gate utilizes three fundamental steps, with qubit states $|0\rangle, |1\rangle$ encoded in the ground-state manifold (see Fig.~\ref{Pulses}). (1) A $\pi$ pulse with Rabi frequency $\Omega_c$ is applied to the first atom, known as the control atom, which excites the $|0 \rangle$ state to a Rydberg state $|c \rangle$. (2) A pulse sequence involving a Rydberg state is applied to the second atom, known as the target atom. Here, we consider a $2 \pi$ pulse with Rabi frequency $\Omega_t$ applied to the $|0 \rangle$ state via the Rydberg state $|t\rangle$ (usually, $|t \rangle = |c\rangle$, but this is not necessary). (3) A $-\pi$ pulse with Rabi frequency $\Omega_c$ is applied to the control atom, returning the Rydberg state to the $|0\rangle$ state.  When the qubits are in the $|1 0 \rangle$ state, they pick up a minus sign due to the $2 \pi$ pulse. Otherwise, the state is left unchanged.

By applying a Pauli-X gate to the target qubit before and after the pulse sequence, this realizes the controlled Z gate (CZ gate), which applies a Pauli-Z gate to the target qubit when the first qubit is in the $|1\rangle$ state. 

Many previous approaches to realizing multi-qubit Rydberg gates rely on the concept of asymmetric Rydberg blockade, in which there is a large separation of scales between different types of Rydberg interactions \cite{Saffman2009,Wu2010a,Wu2017,Isenhower2011,Brion2007a,Su2018}. For example, if the control-control interaction is much smaller than the control-target interaction, then control atoms can blockade target atoms without also blockading other control atoms, which can be used to engineer a multi-control gate. In most cases, asymmetric Rydberg blockade was achieved through the use of strong $1/r^3$ dipole-dipole interactions and weaker $1/r^6$ vdW interactions. However, the dipole-dipole interactions are off-diagonal, which can result in many-body resonances and antiblockade, reducing the gate fidelity \cite{Pohl2009}. 
Moreover, these proposals have been limited to gates involving either many controls or many targets, but not both, which has potential applications for classical verification of quantum computers \cite{Mahadev2018}.

In this Letter, we propose a method for engineering gates involving many control qubits as well as many target qubits. This is accomplished by combining the principles of asymmetric blockade with the conventional two-qubit Rydberg-blockade gate using microwave fields.
The use of microwave fields to modify Rydberg interactions has been considered in a variety of contexts \cite{Bohlouli-Zanjani2007a,Muller2011,Tretyakov2014a, Sevincli2014a, Shi2017,Rao2014,Booth2018a,Petrosyan2014,Wu2017}. We show that by dressing several Rydberg states with strong microwave fields, perfect asymmetric blockade can be realized in which intraspecies (control-control and target-target) Rydberg interactions are negligible while interspecies (control-target) Rydberg interactions are large and nonzero. 
Moreover, the control-target interactions will be \emph{diagonal} dipole-dipole interactions, preventing many-body resonances from playing a role while still utilizing strong dipole-dipole interactions. Since the intraspecies interactions are negligible, the same pulse sequence can be used as in the two-qubit case. This generalizes the CZ gate to a C$_k$Z$^m$ gate with $k$ control qubits and $m$ target qubits (see Fig.~\ref{Pulses}). If all of the control qubits are in the $|1 \rangle$ state, then a Pauli-Z gate is applied to each of the target qubits. Otherwise, the target qubits are unchanged. The same approach can be applied to realize multi-qubit generalization of several other two-qubit Rydberg gates, such as a C$_k U_1 \cdots U_m$ gate which applies a different controlled-unitary to each target qubit \cite{Barenco1995,supplementcite}. Finally, we conclude with a discussion of the performance of these gates compared to other approaches by considering a C$_8$Z$^8$ gate and a simple protocol for creating GHZ states with these gates.

\begin{figure}
\centering
\includegraphics[scale=.31]{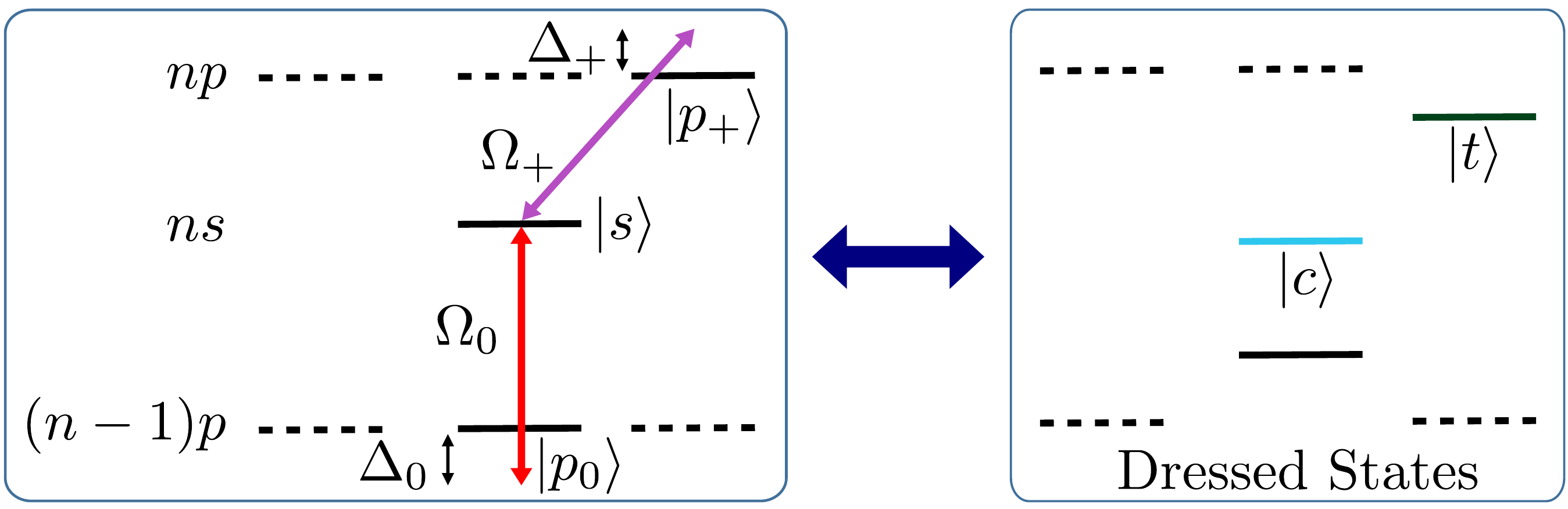}
\caption{Dressing scheme for control and target Rydberg states involving one $s$ state ($L=0$) and two $p$ states ($L=1$), where $n$ denotes the principal quantum number and dotted lines are not involved in the dressing. The $|s\rangle$ state is coupled to the $|p_0\rangle$ state using $\pi$-polarized light with  Rabi frequency $\Omega_0$ and detuning $\Delta_0$. The $|s\rangle$ state is coupled to the $|p_+\rangle$ state using $\sigma^+$-polarized light with Rabi frequency $\Omega_+$ and detuning $\Delta_+$. The right side of the figure illustrates the resulting dressed states $|c\rangle, |t\rangle$, and the third unused dressed state. \label{DressFig}}
\end{figure}

\textit{Microwave dressing}.---
In order to achieve the desired interactions, we consider the dressing scheme shown in Fig.~\ref{DressFig}. This couples a Rydberg $s$ state ($L=0$) to two Rydberg $p$ states ($L=1$) with different principal quantum numbers. Although we study a specific dressing scheme, the only requirement is that one microwave field is $\pi$-polarized while one microwave field is $\sigma$-polarized, which will be used to destructively interfere two interaction terms. Additional drives can be included to provide further tunability. The corresponding Hamiltonian for this dressing, in the rotating frame and under the rotating wave approximation, is
\begin{multline}
\label{dressHam}
H_{mw} = -\Delta_{0} |p_{0} \rangle \langle p_{0} | + \Omega_{0} |s \rangle \langle p_{0}| + \Omega_{0}^* |p_0 \rangle \langle s|\\
-\Delta_{+} |p_{+} \rangle \langle p_{+} | + \Omega_{+} |s \rangle \langle p_{+}| + \Omega_{+}^* |p_+ \rangle \langle s|,
\end{multline}
where $\Delta_{0/+} = \nu_{0/+} - \omega_{0/+}$ denotes the detuning of the drives ($\nu_{0/+}$ and $\omega_{0/+}$ are the drive  and transition frequencies, respectively) and $\Omega_{0/+}$ the Rabi frequency of the drive from the $|s\rangle$ state to the $|p_0\rangle$ and $|p_+\rangle$ states, respectively.

Since the $s$ and $p$ states have different orbital angular momenta, the resultant dressed states experience dipole-dipole interactions. In the rotating frame of both microwave fields, atoms $i$ and $j$ interact via a dipole-dipole interaction
\begin{multline}
V_{dd}^{(i,j)} = \frac{1- 3 \cos^2 \theta_{i j}}{r_{i j}^3} \left(\mu_0^2 | s_i p_{j,0} \rangle \langle p_{i,0} s_j | \right.\\
\left. - \mu_+^2/2 |s_i p_{j,+} \rangle \langle p_{i,+} s_j | \right) + H.c. ,
\label{dressrotDD}
\end{multline}
where $r_{ij}$ is the distance between atoms $i$ and $j$, $\theta_{ij}$ is the angle the displacement vector makes with the quantization axis, and $\mu_0 = \langle p_0| d_0| s \rangle$, $\mu_+ = \langle p_+ |d_+ |s\rangle$ are transition dipole moments, where $d_p = \mathbf{\hat{e}_p} \cdot \mathbf{d}$ is a component of the dipole operator $\mathbf{d}$ and $\mathbf{\hat{e}_0} = \mathbf{\hat{z}},\mathbf{\hat{e}_{\pm}} = \mp (\mathbf{\hat{x}}\pm i \mathbf{\hat{y}})/\sqrt{2}$. 
There are additional interaction terms which do not preserve total $m_L$ (e.g., $|s_i p_{j,+} \rangle \langle p_{i,0} s_j |$) and oscillate with frequencies $2 \nu_+$, $2 \nu_0$, or $\nu_+ \pm \nu_0$ in the rotating frame. When the two $p$ states are from different $p$-state manifolds, these frequencies are all generally large, so the corresponding interactions can be dropped as rapidly oscillating terms in the rotating frame.

\emph{Asymmetric blockade}.---
Next, let us discuss how to design the dressing such that only interspecies interactions are nonzero. Consider a general pair of unnormalized control and target Rydberg states, $|c\rangle$ and $|t\rangle$, which are eigenstates of  Eq.~(\ref{dressHam}), the dressing Hamiltonian
\begin{subequations}
\begin{equation}
|c \rangle \propto |s \rangle + c_0 |p_0 \rangle + c_+ |p_+ \rangle,
\end{equation}
\begin{equation}
|t \rangle \propto |s \rangle + t_0 |p_0 \rangle + t_+ |p_+ \rangle.
\end{equation}
\end{subequations}
In the limit of large drive $\Omega_{0/+} \gg V_{dd}$ \footnote{Sufficiently large microwave Rabi frequencies are achievable since the expectation value of the dipole operator scales like $n^2$.}, the two-atom Rydberg states are product states of the one-atom Rydberg states, i.e., $|cc \rangle, |tt \rangle, |ct \rangle, |t c\rangle$. This holds for $N$-atom Rydberg states as well up to perturbative corrections, which are captured by vdW interactions.
In this basis, the intraspecies interactions for $|c\rangle$ and $|t\rangle$ are
\begin{subequations}
\label{intraint}
\begin{equation}
V_{cc} = \langle cc | V_{dd}| cc \rangle \propto |c_0|^2 \mu_0^2 - |c_+|^2 \mu_+^2/2,
\end{equation}
\begin{equation}
V_{tt} = \langle tt | V_{dd}|tt\rangle \propto |t_0|^2 \mu_0^2 - |t_+|^2 \mu_+^2/2,
\end{equation}
\end{subequations}
where the atom indices $i,j$ have been dropped. From this, we see that while it is not possible to nullify the intraspecies interactions using only a single $p$ state, by dressing an $s$ state with two $p$ states, it is possible to make the two resultant interaction terms destructively interfere due to a difference in sign. This is the origin of the requirement that both $\pi$- and $\sigma$-polarized drives are needed. Thus by fixing $|c_+|^2 = 2 M^2 |c_0|^2$ and $|t_+|^2 = 2 M^2 |t_0|^2$ where $M = \mu_0/\mu_+$, the intraspecies interactions are 0. Although these two constraints are the same for both states, this does not require $|c \rangle = |t\rangle$ because the phases and magnitudes of the coefficients for the two states can be different.

We  must also consider the off-diagonal interactions between $|c\rangle$ and $|t\rangle$. 
The strength of the only resonant off-diagonals term is related to the two intraspecies interactions
$
\langle c t | V_{dd}| t c \rangle \propto  \mathcal{N}_c^4V_{cc} + \mathcal{N}_t^4 V_{tt}
$, where $\mathcal{N}_c, \mathcal{N}_t$ are state normalization factors.
As a result, this interaction is zero when the intraspecies interactions are zero. The remaining off-diagonal terms, such as those proportional to $|cc \rangle \langle tt|$, need not be reduced as long as they are sufficiently off-resonant.

Since the interspecies interaction is the source of Rydberg blockade in the gate, it must be large.
This interaction is
\begin{equation}
V_{ct} = \langle c t| V_{dd} |c t \rangle \propto (c_0 t_0^* + c_0^* t_0) \mu_0^2-(c_+ t_+^* + c_+^* t_+) \mu_+^2/2.
\end{equation}
Although this equation is similar to Eq.~(\ref{intraint}), it differs in that the phases of the coefficients matter. The phases of $c_0, c_+$ can be absorbed into $|p_0\rangle, |p_+\rangle$, leaving $c_0, c_+$ positive with only the phases of $t_0, t_+$ free. The intraspecies interaction is maximized when $t_0, t_+$ are real and have opposite signs. In contrast, it is minimized to zero when they have the same phase.

Additionally, we assume that $|c\rangle$ and $|t\rangle$ come from the same drives, which are applied globally to all atoms. (The case of different drives is discussed in the Supplement \cite{supplementcite}.) This enforces the constraint
\begin{equation}
\langle c| t \rangle \propto 1 + t_0 c_0^* + t_+ c_+^* = 0.
\end{equation}
Taking $c_+ = \sqrt{2} M c_0$ and $t_+ = - \sqrt{2} M t_0$ for real $t_0, c_0$, this becomes $1 + t_0 c_0(1 - 2 M^2) = 0$, from which we find the relation 
\begin{equation}
t_0 = \frac{1}{(2M^2-1)c_0}.
\end{equation}
Thus as long as $M^2 \neq 1/2$, it is possible to realize both dressed states with the same drives. The values of $\Omega_{0/+}, \Delta_{0/+}$ may be determined, up to an overall energy scale, by requiring that both states are eigenvectors of the Hamiltonian for the two drives. The maximum interspecies interaction under this constraint is
\begin{subequations}
\begin{equation}
V_{ct} = \min \left(\frac{\mu_0^2}{\mu_+^2/2},\frac{\mu_+^2/2}{\mu_0^2} \right)(\mu_0^2-\mu_+^2/2),
\end{equation}
\begin{equation}
c_0^{max} = |2 M^2 - 1|^{-\frac{1}{2}},
\end{equation}
\end{subequations}
i.e., the minimum ratio of the two undressed interaction strengths times the difference of the two undressed interaction strengths. We use $c_0^{max}$ to denote the value of $c_0$ which realizes this interaction. The $\min$ function reflects the fact that the larger of the two undressed dipole-dipole interactions will set the overall scale of the interaction. 
Near this maximal interaction strength, the light shifts for $|c\rangle$ and $|t\rangle$ become degenerate, precluding $\pi$ pulses which excite only one or the other and violating the assumption that several off-diagonal interactions are off-resonant. 
To avoid these two issues, the dressing fields should be chosen such that $c_0 = \alpha c_0^{max}$ for $\alpha \neq 1$, removing this degeneracy. While this change in the dressing reduces the interspecies interaction strength, the resulting interspecies interaction remains comparable to the maximal interspecies interaction.

\begin{figure*}
\includegraphics[trim={0 .8cm 0 .7cm},clip,scale=.5]{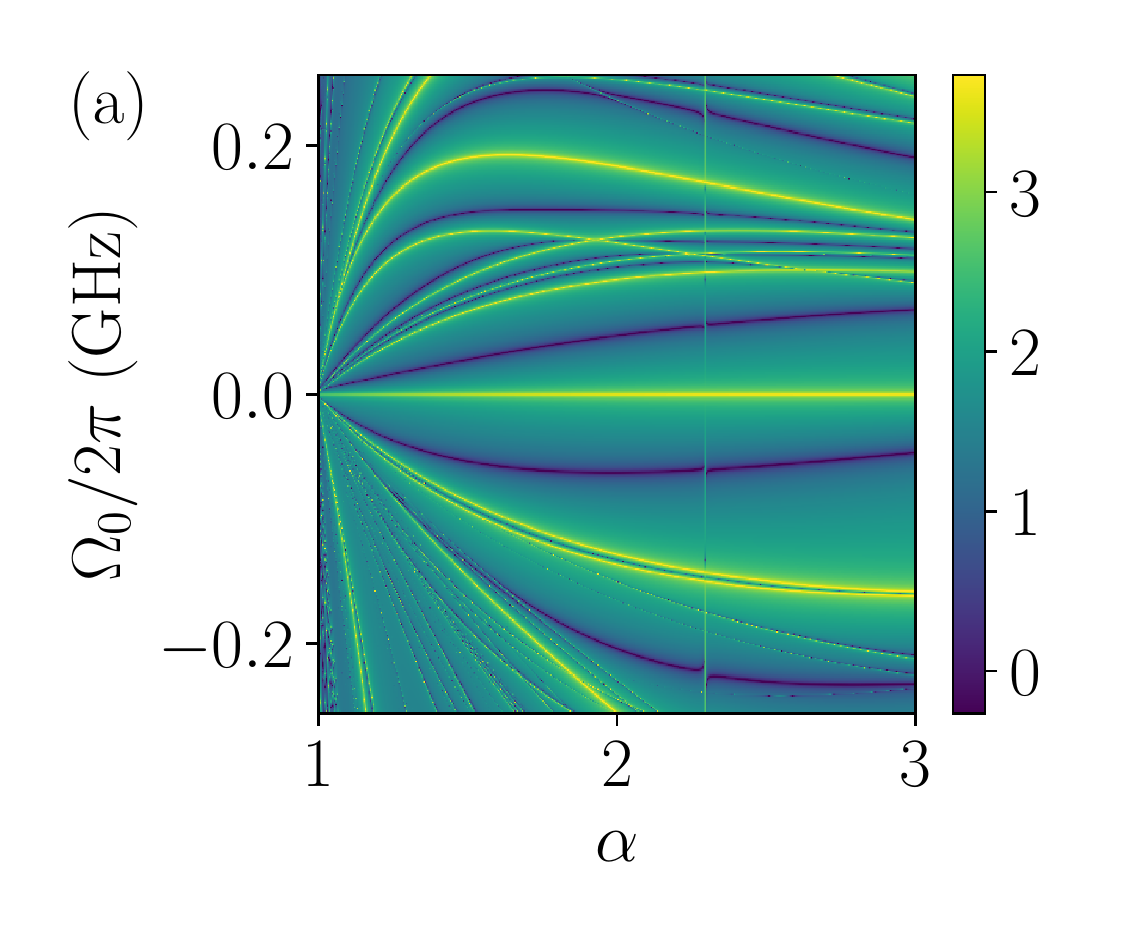}
\includegraphics[trim={0 .8cm 0 .7cm},clip,scale=.5]{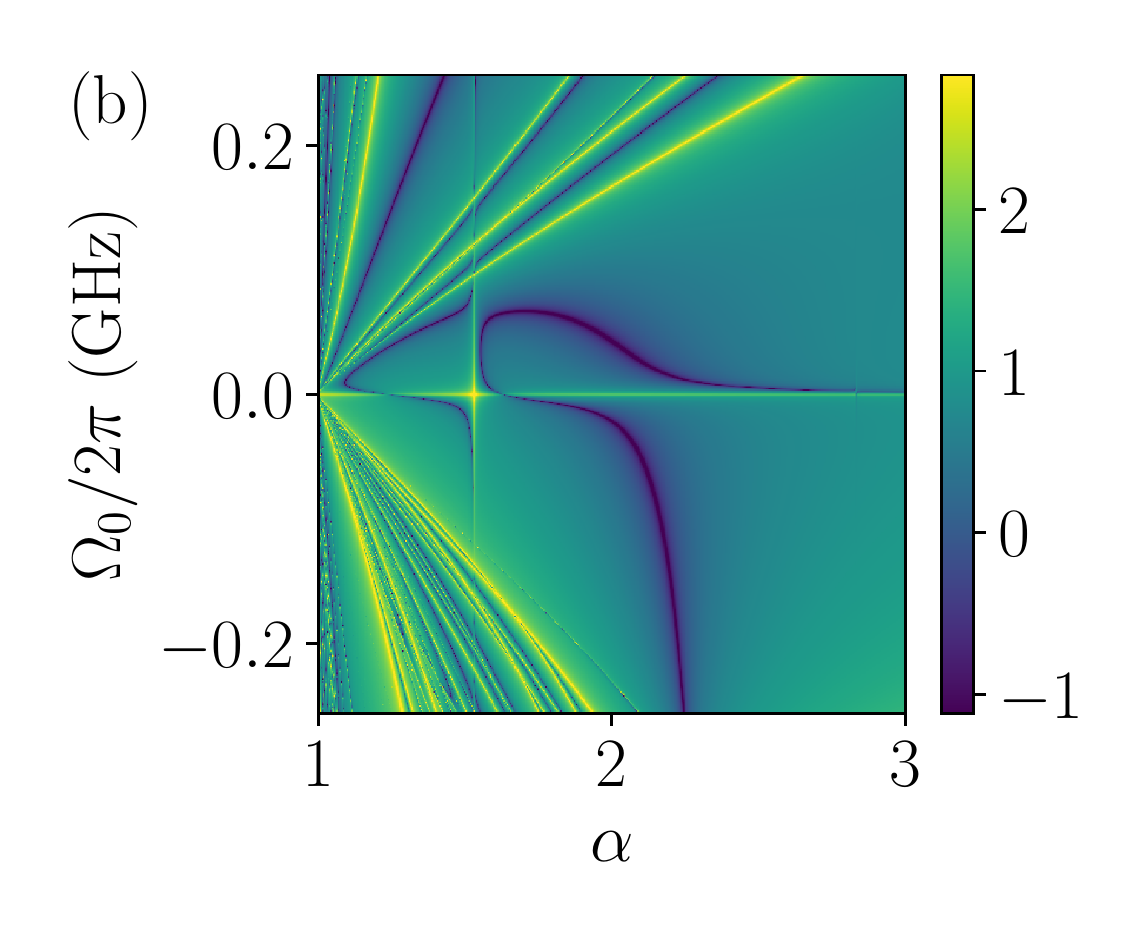}
\includegraphics[trim={0 .8cm 0 .7cm},clip,scale=.5]{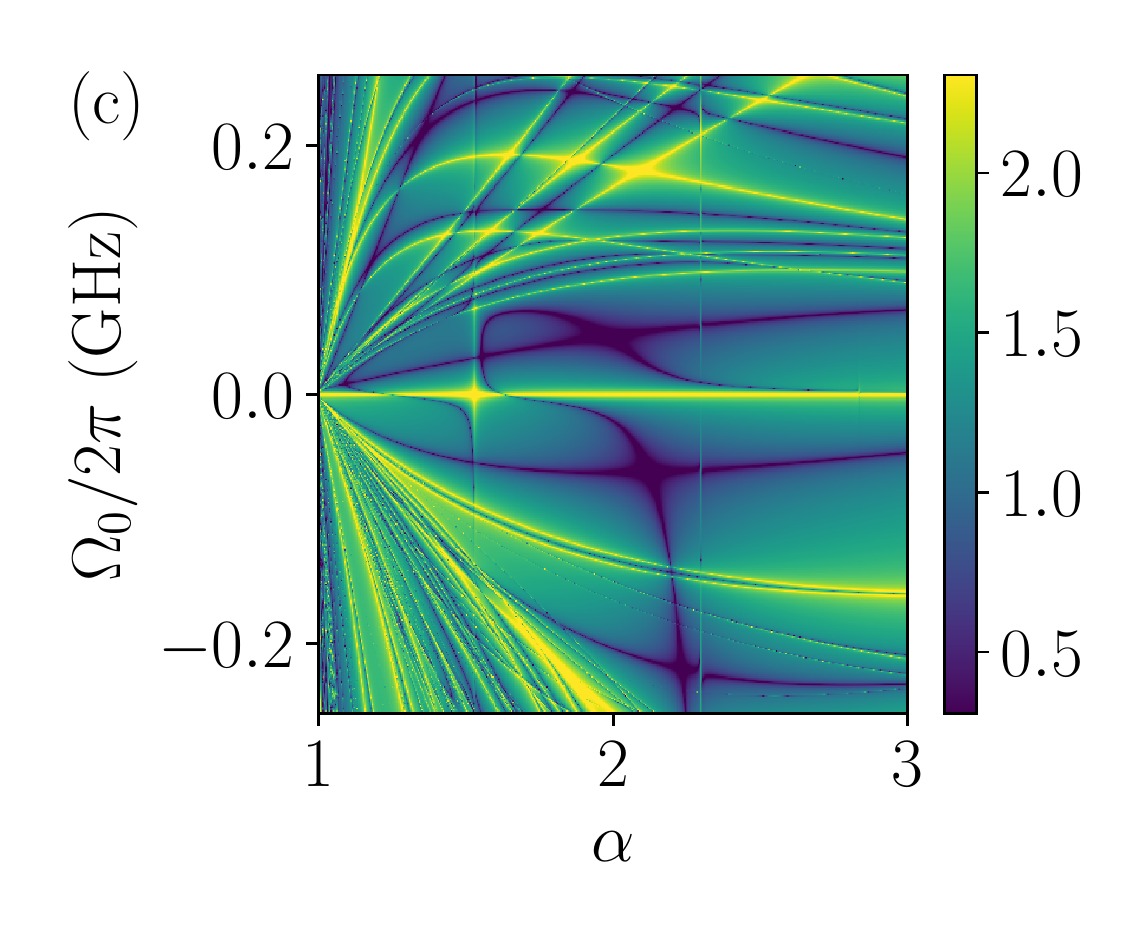}
\caption{Example of vdW nullification for $|s \rangle = |n=60, L=0, J = 1/2, m_J = -1/2 \rangle, |p_0 \rangle = |n=59, L=1, J = 1/2, m_J = -1/2 \rangle, |p_+ \rangle = |n=60, L=1, J = 1/2, m_J = 1/2 \rangle$ of $^{87}$Rb for $\theta = \pi/2$ \cite{supplementcite,Weber2017}. Each plot depicts the behavior of (a) $\log_{10}|C_6^{(c)}/2 \pi|$, (b) $\log_{10}|C_6^{(t)}/2 \pi|$, and (c) $\frac{1}{2} \log_{10}|C_6^{(c)} C_6^{(t)}/(2 \pi)^2 |$ as a function of $\alpha, \Omega_0$, where $C_6$ values are in units of GHz $\mu$m$^6$. 
The lines where the value of $C_6$ appears to diverge corresponds to the presence of a resonant interaction which cannot be treated perturbatively.
The resonances at fixed $\alpha$ arise due to degeneracies in the states coupled by the drives. Another resonance occurs for $\Omega_0 = 0$ due to resonant dipole-dipole interactions involving $s, p$ states in the same manifolds as $|s\rangle, |p_0\rangle, |p_+\rangle$. There are a variety of points where the vdW interactions become negligible and a perturbative approach is valid. One such point occurs at $\alpha \approx 2.2$, $\Omega_0/2 \pi \approx -.22$ GHz, where $\Omega_+ \approx \Omega_0$. The lifetimes of $|c\rangle$ and $|t\rangle$ at this point are $\tau_c = .4$ ms and $\tau_t = .48$ ms, respectively \cite{Sibalic2017}. \label{NullFig}}
\end{figure*}

\textit{Nullifying vdW interactions}.---
Since we have successfully eliminated the intraspecies dipole-dipole interactions for $|c\rangle$ and $|t\rangle$, intraspecies vdW interactions are relevant. While the dipole-dipole interactions are much larger than the vdW interactions for the same atomic separation, it is important to compare intraspecies interactions at short distances to interspecies interactions at long distances. 
The target-target vdW interaction is particularly important, as $\Omega_t$ must be simultaneously stronger than the vdW interaction and weaker than the blockade interaction $V_{ct}$. In contrast, $\Omega_c$ is not limited by $V_{ct}$.

The intraspecies vdW interactions take the form
\begin{equation}
V_{vdW}^{(i,j)} = - \frac{C_6^{(c)}}{r_{ij}^6} | c_i c_j \rangle \langle c_i c_j|- \frac{C_6^{(t)}}{r_{ij}^6} | t_i t_j \rangle \langle t_i t_j|,
\end{equation}
where $C_6^{(c)}, C_6^{(t)}$ denote the strength of the intra-species vdW interactions for $|c\rangle$ and $|t\rangle$, respectively. These strengths are a result of second-order non-degenerate perturbation theory. Since the off-resonant coupling strengths and energy differences are dependent on the dressing, the strength of the vdW interactions changes as a function of the dressing, making $C_6^{(c)}, C_6^{(t)}$ tunable. Given the constraints on the dressing, there are two degrees of freedom allowing this tunability. The first is simply the overall energy scale of the dressing fields. By varying $H_{mw}$ by a constant factor, the dressed states remain the same while the light shifts change, modifying the off-resonant energy differences in the perturbative calculation of $C_6$. The second degree of freedom is encoded in $\alpha$, representing the fact that we have four degrees of freedom in the dressed states ($c_0, c_+,t_0,t_+$) and three constrains (orthogonality and no intraspecies interactions).

Since there are two tunable parameters in the dressing and two $C_6$ coefficients, this opens the possibility of tuning both values of $C_6$ to zero. As we illustrate in Fig.~\ref{NullFig} for a particular choice of states, this is indeed possible. For both $|c\rangle$ and $|t\rangle$, there are lines where their individual vdW interactions become zero, and these lines can intersect. The presence of these lines can be understood by considering the existence of two-atom resonances, which appear as divergences and correspond to a breakdown in our non-degenerate perturbative approach. At one of these resonances, the energy difference passes through zero and $C_6$ changes signs, leading to zero crossings when there are multiple resonances. Although zero crossings are often near a resonance, there are several parameter regimes where this is not the case and nearby resonances are not an issue. These zero crossings can be identified experimentally via two-atom blockade and antiblockade experiments.

\textit{Gate performance}.---
There are several factors to take into account when considering the implementation of multi-qubit gates based on asymmetric blockade. These fall into two different categories: the validity of the dressed state picture and errors in the gate performance. Below, we discuss the constraints that arise from both.

There are two primary requirements for the dressed state picture to be valid: (i) the Rabi frequencies of the dressing fields must be strong compared to the dipole-dipole interactions and (ii) the dressing fields must not couple the dressed states to any additional Rydberg states. While (ii) depends on the exact details of the level structure and dressing, it roughly corresponds to $C_3/l_{dd}^3 \ll \Omega_{mw} \ll \delta$, where $l_{dd}$ is the smallest dipole-dipole interaction distance, $\Omega_{mw}$ defines the scale of the  microwave dressing, and $\delta$ is on the order of a two-atom energy defect (the energy difference between a pair state composed of any two of the $|s\rangle, |p_0\rangle,$ and $|p_+\rangle$ states and a nearby two-atom Rydberg state \cite{Saffman2010}). Importantly, $\Omega_{mw} \ll \delta$ is only necessary to confine the dressing to three states and simplify the resulting analysis. Thus stronger control-target interactions can be realized at the cost of more complicated analysis of the dressed states.

There are three primary sources of error for the gate: dissipation, imperfect blockade, and nonzero vdW interactions. For a square $2 \pi$ pulse or two square $\pi$ pulses, the probability of decay for a single Rydberg state is $\epsilon_\gamma = \frac{\pi/2}{\Omega_g \tau}$, where $\Omega_g$ corresponds to the Rabi frequency used and $\tau$ corresponds to the the lifetime of the Rydberg state. The error due to imperfect blockade is approximately $\epsilon_\pi \approx (2 \Omega_g^{(t)}/V_b)^2$, while we approximate the error due to nonzero vdW interactions as $\epsilon_{vdW} \approx (V_{vdW}/(2\Omega_g))^2$, where $V_{vdW}$ is the total blockade strength of the vdW interactions. In general, the errors from the $\pi$ pulses for the control qubits can be neglected because the Rabi frequency is not constrained by the control-target interactions, so the Rabi frequency can be made very large.

In order to investigate the performance of these gates, we consider two scenarios. In the first, we consider a C$_8$Z$^8$ gate on a $4 \times 4$ lattice in which control and target atoms occupy different square sublattices (see Fig.~\ref{Pulses}) and the initial state is $|+ \rangle^{\otimes 16}$, where $|+ \rangle = (|0 \rangle + |1 \rangle)/\sqrt{2}$, to capture the average error. Using the dressing at the point discussed in Fig.~\ref{NullFig}, the maximal nearest-neighbor control-target interaction is $2 \pi \times 2.7$ MHz, which is 80 times smaller than the smallest microwave Rabi frequency. A factor of 10 is to ensure the microwave fields are stronger than the undressed dipole-dipole interactions while a factor of 8 is due to a reduction in the dressed dipole-dipole interactions compared to the undressed dipole-dipole interactions. Optimizing the pulse strengths, we find an error of $10\%$.

In the second scenario, we use these gates to create 13- and 25-atom GHZ states using two or three steps, respectively. This is achieved by using a protocol inspired by Ref.~\cite{Eldredge2017}. Additionally, we utilize C$_k$NOT$^m$ gates, which can be realized by applying single-qubit Hadamard gates to the target qubits before and after the C$_k$Z$^m$ gate. Initially, all qubits in a square lattice are in $|0\rangle$ except for one, which starts in $(|0\rangle + |1\rangle)/\sqrt{2}$. Using the latter qubit as a control and its nearest neighbors as targets, a CNOT$^4$ gate is applied, creating a 5-atom GHZ state. This process can be repeated using the boundary as controls and their outer nearest neighbors as targets, quickly increasing the size of the GHZ state at each step. The 13-atom GHZ state has an error of 4.5\% while the 25-atom GHZ state has an error of 7.8\% \cite{supplementcite}. In comparison, Ref.~\cite{Saffman2009} predicts 16\% error for an 8-atom GHZ state via asymmetric blockade. Although two-qubit gates with a theoretical minimal error of .3\% have comparable errors (3.6\% and 7.2\%), they require 12 and 24 gates, respectively, as well as much larger Rabi frequencies \cite{Saffman2005}.

\textit{Outlook}.---
We have presented a protocol which uses microwave-dressed Rydberg states to realize multi-qubit gates involving multiple control qubits and multiple target qubits. These gates can be used to simplify quantum protocols, greatly reducing the number of gates needed. While this helps reduce the need for fault-tolerant error correction, understanding how to realize fault-tolerance for complicated multi-qubit gates remains an important direction \cite{Preskill1998,Aharonov2008}. Although we have considered only the situation in which three Rydberg states are dressed together, these principles can be generalized to situations in which additional Rydberg states are coupled via microwave dressing, providing more tunability. Moreover, the application of strong microwave fields provides a new approach to realizing novel, tunable interactions for quantum simulation, and could also be used for nondestructive cooling by engineering state-insensitive interactions \cite{Belyansky2019} or monitoring quantum simulators with quantum non-demolition couplings \cite{Vasilyev2020}. 
Similarly, it is worth exploring ways to realize more general multi-qubit gates without two-qubit counterparts. For example, more general forms of controlled-unitary gates and controlled Hamiltonian evolution, which has potential applications in anyonic interferometry \cite{Jiang2008}, measuring quantum information scrambling \cite{Swingle2016}, quantum phase estimation \cite{OBrien2019}, and quantum metrology with indefinite causal order \cite{Zhao2020a}, and which also has close connections to the central spin model \cite{Yao2006}. Additionally, these methods have potential applications in speeding up state transfer and the preparation time of MERA (multiscale entanglement renormalization ansatz) using the long-range $1/r^3$ interactions \cite{Eldredge2017,Tran2020}. 
Both dipole-dipole interaction gates \cite{Yelin2006, Ni2018,Hudson2018} and microwave-dressed dipole-dipole interactions \cite{Buchler2007,Gorshkov2011} have been discussed in the context of polar molecules. Therefore, the ideas presented in this Letter can be applied to polar molecules or other systems with dipole-dipole interactions. Similarly, these methods may have potential applications in systems with magnetic dipole-dipole interactions, such as nitrogen-vacancy centers in diamond \cite{Bermudez2011,Dolde2013} and magnetic atoms \cite{Griesmaier2005,Lu2011,Aikawa2012}.

\begin{acknowledgments}
We thank M.\ Lukin, G.\ Alagic, A.\ Childs, O.\ Shtanko, L.P.\ Garc\'ia-Pintos, A.\ Deshpande, and M.\ Tran for discussions. J.T.Y., P.B., R.B., and A.V.G.\ acknowledge support by AFOSR, AFOSR MURI, DoE ASCR Quantum Testbed Pathfinder program (award No.\ {DE-SC0019040}), DoE BES Materials and Chemical Sciences Research for Quantum Information Science program (award No.\ {DE-SC0019449}), DoE ASCR FAR-QC (award No.\ {DE-SC0020312}), NSF PFCQC program, ARO MURI, ARL CDQI,  and NSF PFC at JQI. A.M.K.\ acknowledges support by AFOSR and NIST. R.B.\ acknowledges support by NSERC and FRQNT of Canada.
\end{acknowledgments}

\end{document}


\title{Supplemental Material: Asymmetric blockade and multi-qubit gates via dipole-dipole interactions}	

\author{Jeremy T. Young}
\affiliation{Joint Quantum Institute, NIST/University of Maryland, College Park, Maryland 20742 USA}

\author{Przemyslaw Bienias}
\affiliation{Joint Quantum Institute, NIST/University of Maryland, College Park, Maryland 20742 USA}

\author{Ron Belyansky}
\affiliation{Joint Quantum Institute, NIST/University of Maryland, College Park, Maryland 20742 USA}

\author{Adam M. Kaufman}	

\affiliation{JILA, University of Colorado and National Institute of Standards and Technology, and Department of Physics, University of Colorado, Boulder, Colorado 80309, USA}

\author{Alexey V. Gorshkov}
\affiliation{Joint Quantum Institute, NIST/University of Maryland, College Park, Maryland 20742 USA}
\affiliation{Joint Center for Quantum Information and Computer Science, NIST/University of Maryland, College Park, Maryland 20742 USA}

\date{\today}

\maketitle
\onecolumngrid

\renewcommand{\theequation}{S\arabic{equation}}
\renewcommand{\thesubsection}{S\arabic{subsection}}
\renewcommand{\thesubsubsection}{\Alph{subsubsection}}

\renewcommand{\bibnumfmt}[1]{[S#1]}
\renewcommand{\citenumfont}[1]{S#1} 

\pagenumbering{arabic}

\makeatletter
\renewcommand{\thefigure}{S\@arabic\c@figure}
\renewcommand \thetable{S\@arabic\c@table}

This supplemental material is organized as follows: in Sec.~\ref{suppcU}, we illustrate how to use the C$_k$Z$^m$ gates from the main text to realize generic C$_kU_1 \cdots U_m$ gates, in which an arbitrary unitary $U_i$ gate is applied to the $i$th qubit. In Sec.~\ref{ctdriveSupp}, we discuss how control and target atoms can be dressed with different fields. In Sec.~\ref{vdWsupp}, we present the method used for calculating perturbatively the van der Waals interactions for the dressed states. In Sec.~\ref{GHZsupp}, we present the details of the protocol used in the main text to prepare large GHZ states using the multi-qubit gates.

\section{Realizing a controlled unitary gate}
\label{suppcU}
In this section, we demonstrate how the multi-qubit C$_k$Z$^m$ gates in the main text can be generalized to generic C$_kU_1 \cdots U_m$ gates, in which an arbitrary unitary $U_i$ gate is applied to the $i$th qubit. This is achieved in a manner that is similar to the generalization for two-qubit gates and relies on single-qubit gates \cite{Barenco1995}. In particular, we rely on the fact that a single-qubit unitary can be written in the form $U = e^{i \delta} W$, where $W \in SU(2)$. Additionally, we take advantage of the fact that there exist matrices $A,B,C \in SU(2)$ such that $A B C = I$ and $W =A Z B Z C$ \cite{Barenco1995}. 

The latter of these two identities implies a simple way to realize a two-qubit C$W$ gate: apply $A$, CZ, $B$, CZ, $C$, where $A,B,C$ are applied to the target qubit. The generalization to multiple controls and targets is straightforward. For each target qubit, we choose $A_i, B_i, C_i$ such that $W_i$ is applied to that qubit.

\begin{figure}[h]
    \centering
    \includegraphics[scale=.33]{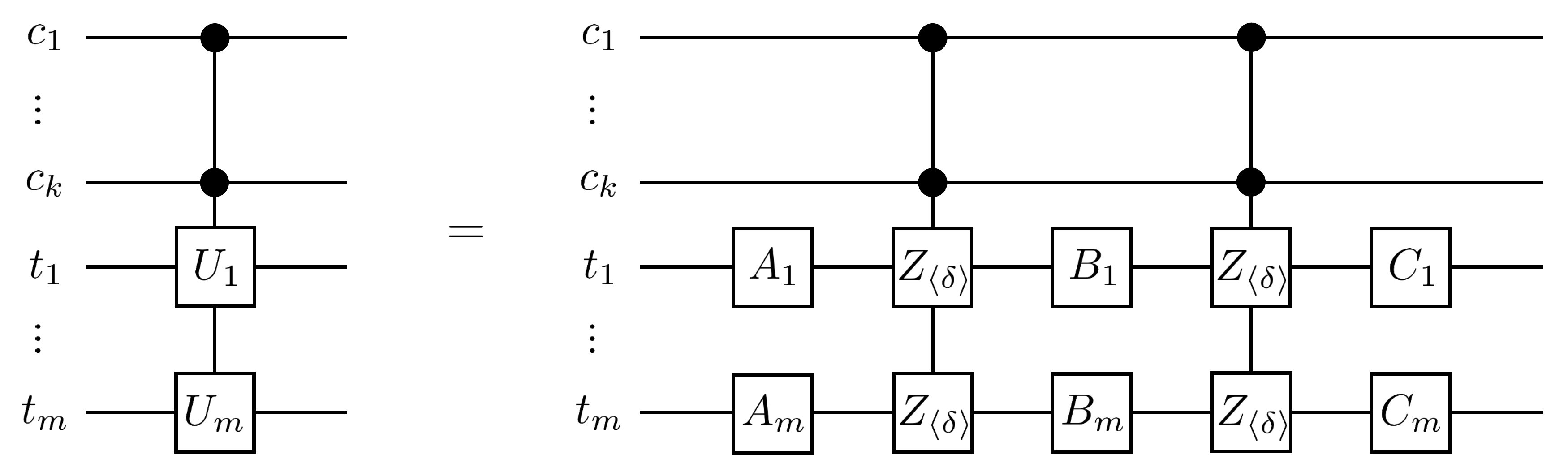}
    \caption{Generalization of C$_k$Z$^m$ gates to C$_kU_1 \cdots U_m$ for general unitaries $U_i$. The control and target qubits are denoted by $c_i$ and $t_i$, respectively. $A_i, B_i, C_i \in SU(2)$ with $A_i B_i C_i = I$, and $Z_{\langle \delta \rangle} = e^{i \langle \delta \rangle/2} Z$. The unitaries applied to each target qubit are $U_i = e^{i \langle \delta \rangle} A_i Z B_i Z C_i$, which is equivalent to applying the unitaries $U_i = e^{i \delta_i} A_i Z B_i Z C_i$, where $ \frac{1}{m}\sum_i \delta_i = \langle \delta \rangle$.}
    \label{cU}
\end{figure}

To fully realize general $U_i$, a phase needs to be applied. This can be achieved through a slight modification of the pulse sequence on the target atoms. In the main text, we assumed that the $2 \pi$ pulse was applied using the same Rabi frequency throughout. If we instead apply two $\pi$ pulses using Rabi frequencies with different phases, the $|1 0 \rangle$ state will pick up an extra phase based on the difference of the Rabi frequency phases, and $|1 0 \rangle \to - e^{i \delta/2} |10\rangle$. By applying an extra pair of $\pi$ pulses from the $|1\rangle$ state of the target qubits to the $|t\rangle$ state, we can realize $|1 1 \rangle \to e^{i \delta/2} |1 1 \rangle$. Applying the necessary Pauli-X gates to the target qubit, the resulting gate applied to the target qubit is $Z_{\delta} \equiv e^{i \delta/2} Z$. Using this controlled gate in place of the CZ gates above, the corresponding unitary is then $U_i = e^{i \delta_i} W_i$. Since the phases for all the $U_i$ add together, we only need to realize a single overall phase $\sum_i \delta_i$. To reduce the local control of the phase needed, we use the same phase $\langle \delta \rangle = \frac{1}{m}\sum_i \delta_i$ for all the target qubits. This gate sequence is illustrated in Fig.~\ref{cU}. 

We could alternatively apply the necessary phase $\sum_i \delta_i$ through just one of the target qubits, while the remaining target qubits contribute no phase. Additionally, in the case where there is only a single control qubit, we only need to apply a phase $\sum_i \delta_i$ to the $|1\rangle$ state of the control qubit via a single-qubit gate. 

\section{Different driving for control and target atoms}
\label{ctdriveSupp}
In this section, we discuss the scenario in which the control and target atoms can be dressed with different fields. This would rely on using dressing fields which have wavelengths which are small compared to the atomic spacing. This could be achieved, for example, via a two-photon process through a low-energy state (e.g.,~a ground state) which couples the $|p_0 \rangle$ and $|p_+\rangle$ states, although large Rabi frequencies will be more difficult to achieve.

For fixed magnitudes of the coefficients of the dressed states, the intraspecies interaction is maximized when $t_0, t_+$ are real and have opposite signs. This can be achieved, for example, by taking $\Omega_{c,0} = \Omega_{t,0}$, $\Omega_{c,+} = - \Omega_{t,+}$, where the first subscript denotes whether the drive is applied to a control ($c$) or target ($t$) atom and the second subscript denotes the polarization of the drive. Taking normalization factors into account, the maximal interaction is given by
\begin{equation}
\langle c t| V_{dd} |c t \rangle = \pm \frac{4 c_0 t_0 \mu_0^2}{\mathcal{N}_0^2 \mathcal{N}_1^2} \frac{1 - 3 \cos^2 \theta}{r^3},
\end{equation} 
where $\mathcal{N}_0^2 = 1+ (1+2 M^2) |c_0|^2$,  $\mathcal{N}_1^2 = 1 + (1+2 M^2)|t_0|^2$ are normalization factors and $M = \mu_0/\mu_+$. The interaction is maximized (in magnitude) when $c_0 = \pm t_0 = \sqrt{1+2M^2}^{-1}$, where it takes the value
\begin{equation}
\langle c t| V_{dd} |c t \rangle = \pm \frac{1}{\mu_0^{-2} + 2 \mu_+^{-2}} \frac{1 - 3 \cos^2 \theta}{r^3},
\end{equation} 
which is the harmonic mean of the the two individual dipole-dipole interactions. This is generally larger than can be achieved when the atoms are dressed with the same drives. Off-diagonal interactions need not be considered in general, as the different drives will result in different light shifts, so the off-diagonal interactions will be off-resonant.

\section{Calculation of van der Waals interactions}
\label{vdWsupp}
In this section, we present the method used for calculating van der Waals interactions for the dressed states. We assume that the atomic separation is sufficiently large that $C_6$ can be determined via second-order perturbation theory. Due to the degeneracy of the states used in the dressing with undressed states and the fact that the microwave Rabi frequencies are large compared to the dipole-dipole interactions, it is important to take into account light shifts due to the dressing. In the rotating frame, the Hamiltonian of the system is given by
\begin{equation}
H = E_c |c \rangle \langle c| + E_t |t \rangle \langle t| + E_3 |3 \rangle \langle 3| + \sum_u E_u |u \rangle \langle u| + V_1 + V_2 + V_3,
\end{equation}
where $|3 \rangle$ is the third dressed state which is not involved in the Rydberg gate, $|u \rangle$ are undressed Rydberg states, $E_\mu$ is the energy (including light shifts) of state $|\mu \rangle$ in the rotating frame, and there are three types of dipole-dipole interaction terms
\begin{subequations}
\begin{equation}
    V_1 = \sum_{u, u'} \sum_{\sigma, \sigma'} V_{w_\sigma u}^{w_{\sigma'} u'} e^{i (\nu_\sigma + \nu_{\sigma'}) t} |w_\sigma w_{\sigma '} \rangle \langle u u'| + H.c.,
\end{equation}
\begin{equation}
    V_2 = \sum_{d_1,d_2,d_3,d_4} V_{d_1 d_3}^{d_2 d_4} |d_1 d_2 \rangle \langle d_3 d_4| + H.c.,
\end{equation}
\begin{equation}
    V_3 = \sum_{d_1,d_2,d_3,u} V_{d_1 d_3}^{d_2 u}(t)  (|d_1 d_2 \rangle \langle d_3 u|+|d_2 d_1 \rangle\langle u' d_3|) + H.c.,
\end{equation}
\end{subequations}
where $\sigma = \{s, 0, +\}, \{w_s,w_0,w_+\} \equiv \{s, p_0, p_+\}, d_i \in \{c, t, 3\}$, and $V^{\psi' \phi'}_{\psi \phi} = \langle \psi \psi' | V_{dd} | \phi \phi' \rangle$, where $V_{dd}$ is the dipole-dipole interaction. Thus $V_1$ corresponds to undressed intermediate states, $V_2$ to dressed intermediate states, and $V_3$ to a mix of dressed and undressed intermediate states. The mixed term $V_3$ must be put in the basis of the dressed states in order to properly identify the energies of the intermediate states. All three are from dipole-dipole interactions in the necessary basis. Due to the rotating frames, $V_1$ possess rotating terms. Similarly, $V_3$ also possesses rotating terms which are more complicated due to the change of basis. Defining $|w_\sigma \rangle = \sum_i R_{\sigma i} |d^i \rangle$ with $\{d^1, d^2, d^3\} \equiv \{c, t, 3 \}$, then the $V_3$ terms can be written
\begin{equation}
    V_{d_1 d_3}^{d_2 u}(t) = \sum_{\sigma, \sigma', \sigma''} V_{w_\sigma w_{\sigma''}}^{w_{\sigma'} u} e^{i (\nu_\sigma + \nu_{\sigma'} - \nu_{\sigma''})t} R_{\sigma 1} R_{\sigma' 2} R_{\sigma '' 3}^*,
\end{equation}
which may possess multiple terms rotating with different frequencies.

Due to the presence of multiple rotating frames, there is no single rotating frame which removes all time dependence, which is necessary to apply time-independent perturbation theory. Instead, we apply a Floquet approach and expand the states in a quasi-energy series
\begin{equation}
|u u' \rangle = \sum_{n, m} e^{i n \nu_m t |u u{'}^m \rangle},
\end{equation} 
where $n$ labels the harmonic and $m$ labels the different rotating frames needed for a given state. Practically speaking, this has the effect of shifting the energy defect for a given rotating term by $n \nu_m$. With this in mind, we can write the perturbative corrections from each of the three interaction terms
\begin{subequations}
\label{vdWterms}
\begin{equation}
    V_{1,cc}^{vdW} =  \sum_{\sigma, \sigma'} \sum_{u,u'} \frac{ |a_\sigma a_{\sigma'} V_{u_\sigma u}^{u_{\sigma'} u'}|^2}{2 E_c + \nu_\sigma + \nu_{\sigma'} - E_{u} - E_{u'}},
    \label{vdW1}
\end{equation}
\begin{equation}
    V_{2,cc}^{vdW} = \sum_{(d_1, d_2) \neq(c,c)} \frac{|V_{d_1 c}^{d_2 c}|^2}{2 E_c - E_{d_1} -E_{d_2}},
    \label{vdW2}
\end{equation}
\begin{equation}
    V_{3,cc}^{vdW} = 2 \sum_{ \sigma, \sigma', \sigma''} \sum_{d, u}  \frac{ |R_{\sigma c} R_{\sigma' c} R_{\sigma'' d}^* V_{w_\sigma w_{\sigma''}}^{w_{\sigma'} u}|^2}{2 E_c + \nu_\sigma + \nu_{\sigma'} - \nu_{\sigma ''} - E_d - E_{u}},
\end{equation}
\end{subequations}
where we have, without loss of generality, focused on the control-control vdW interactions and $a_s, a_\sigma$ are the normalized coefficients of $|c\rangle$. Eqs.~(\ref{vdW1},\ref{vdW2}), are the contributions of the typical $s$ and $p$ state vdW interactions with the effect of the light shifts included. The third term is a new contribution due to the dressing. In all three cases, the effects of the light shifts must be included, as they are needed to tune the vdW interactions to 0. This approach is easily generalized to cases where additional states are coupled due to the drives.

\section{GHZ state preparation}

\label{GHZsupp}

In this section, we present the details of the protocol used to prepare large GHZ states using the multi-qubit gates in the main text. This approach is inspired by the protocol in Ref.~\cite{Eldredge2017}. In order to prepare a GHZ state, we rely on the fact that a controlled NOT (CNOT) gate has the following behavior:
\begin{equation}
\text{CNOT} \left(\frac{|0 0 \rangle + |1 0 \rangle}{\sqrt{2}}\right) = \frac{|0 0 \rangle + |11 \rangle}{\sqrt{2}}.
\end{equation}
By using any qubit part of the GHZ state as a control qubit and a target qubit in the $|0\rangle$ state, the size of the GHZ state can be sequentially increased. By using the multi-qubit gates developed in the main text, many qubits can be incorporated into the GHZ state in a single step. Although the gate in the main text is a C$_k$Z$^m$ gate, a C$_k$NOT$^m$ can be realized either via a modification to the pulse sequence or by applying single-qubit Hadamard gates to the target qubits before and after the C$_k$Z$^m$ gate; we consider the latter implementation.

\begin{figure}[h]
\subfloat[]{
\includegraphics[scale=.4]{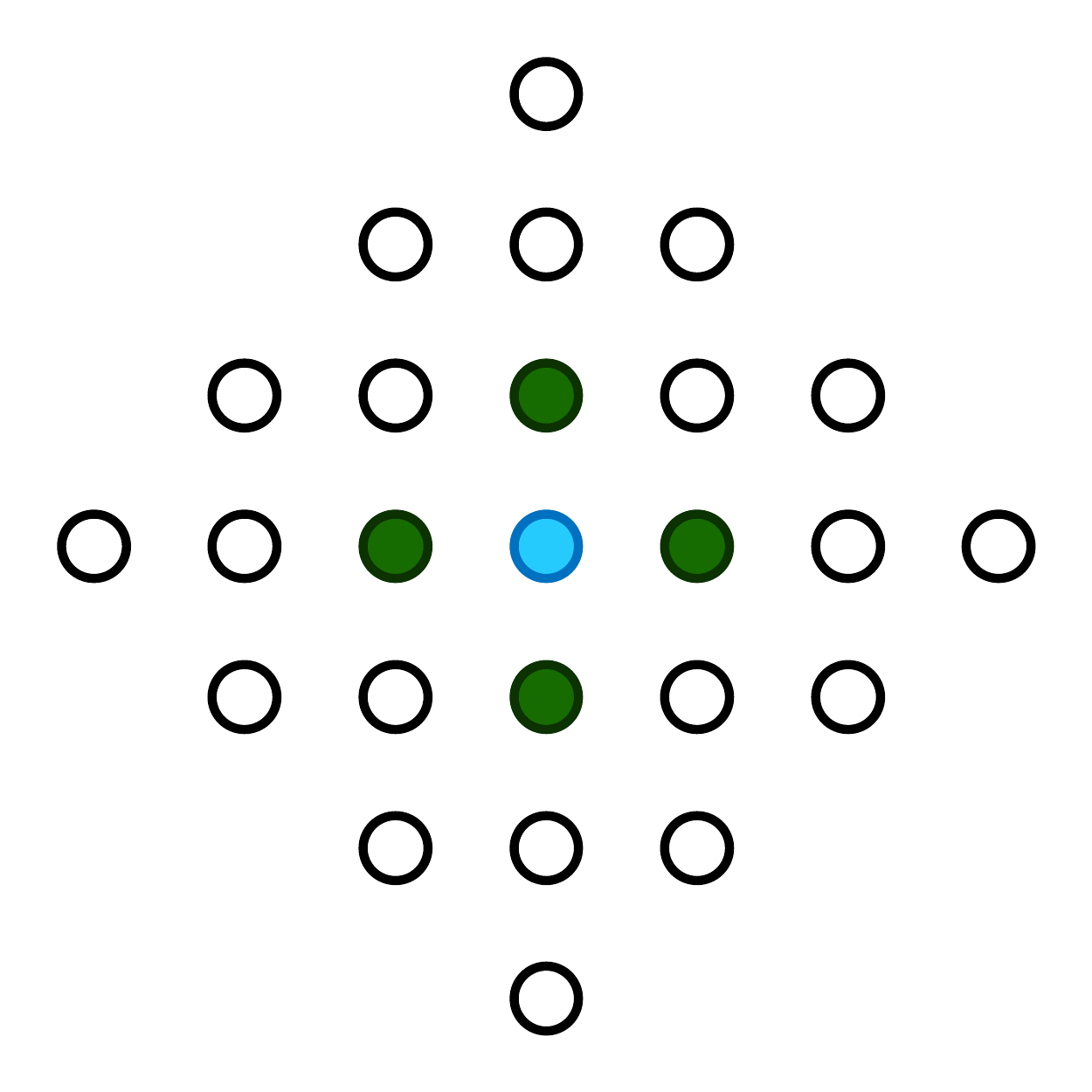}
}
\qquad
\subfloat[]{
\includegraphics[scale=.4]{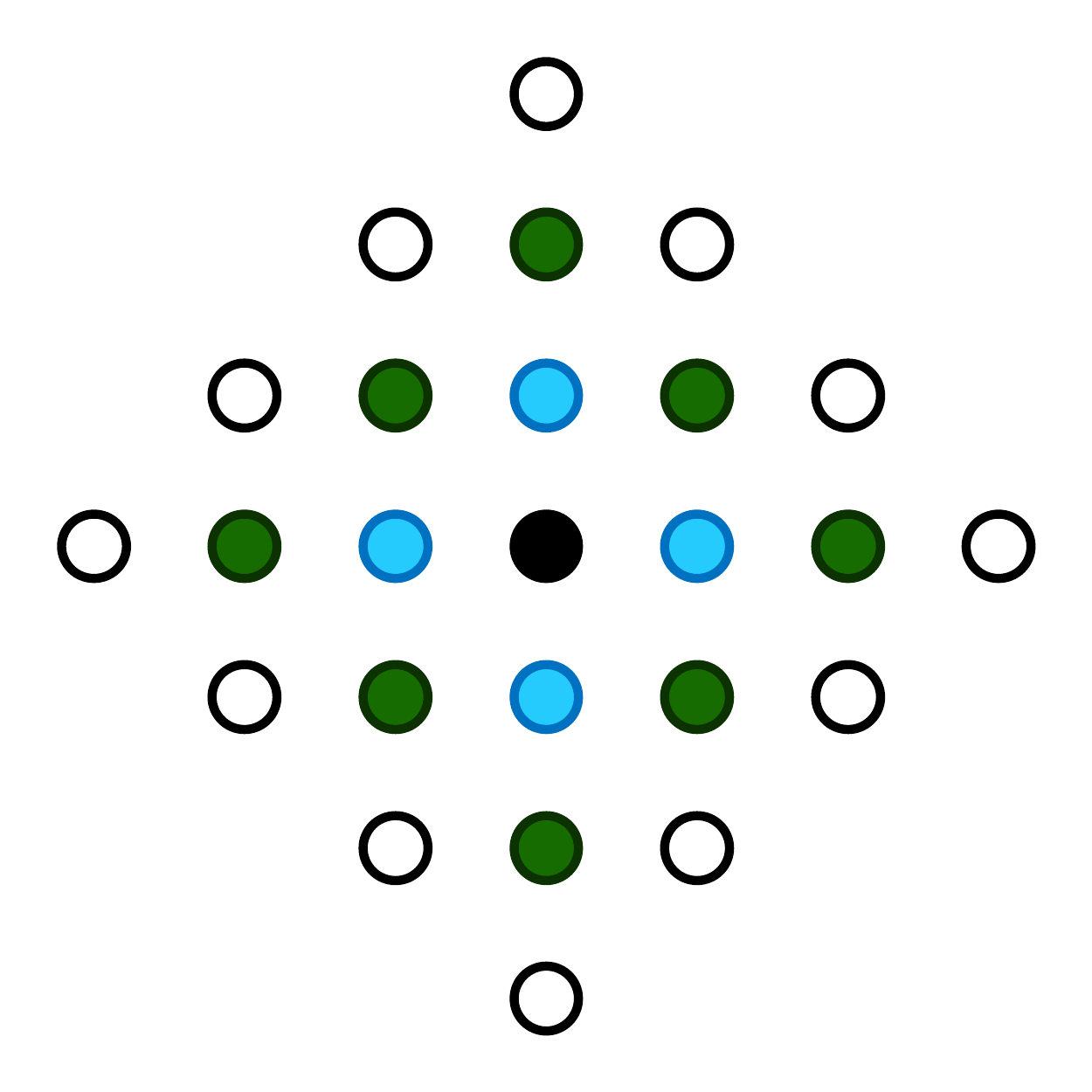}
}
\qquad
\subfloat[]{
\includegraphics[scale=.4]{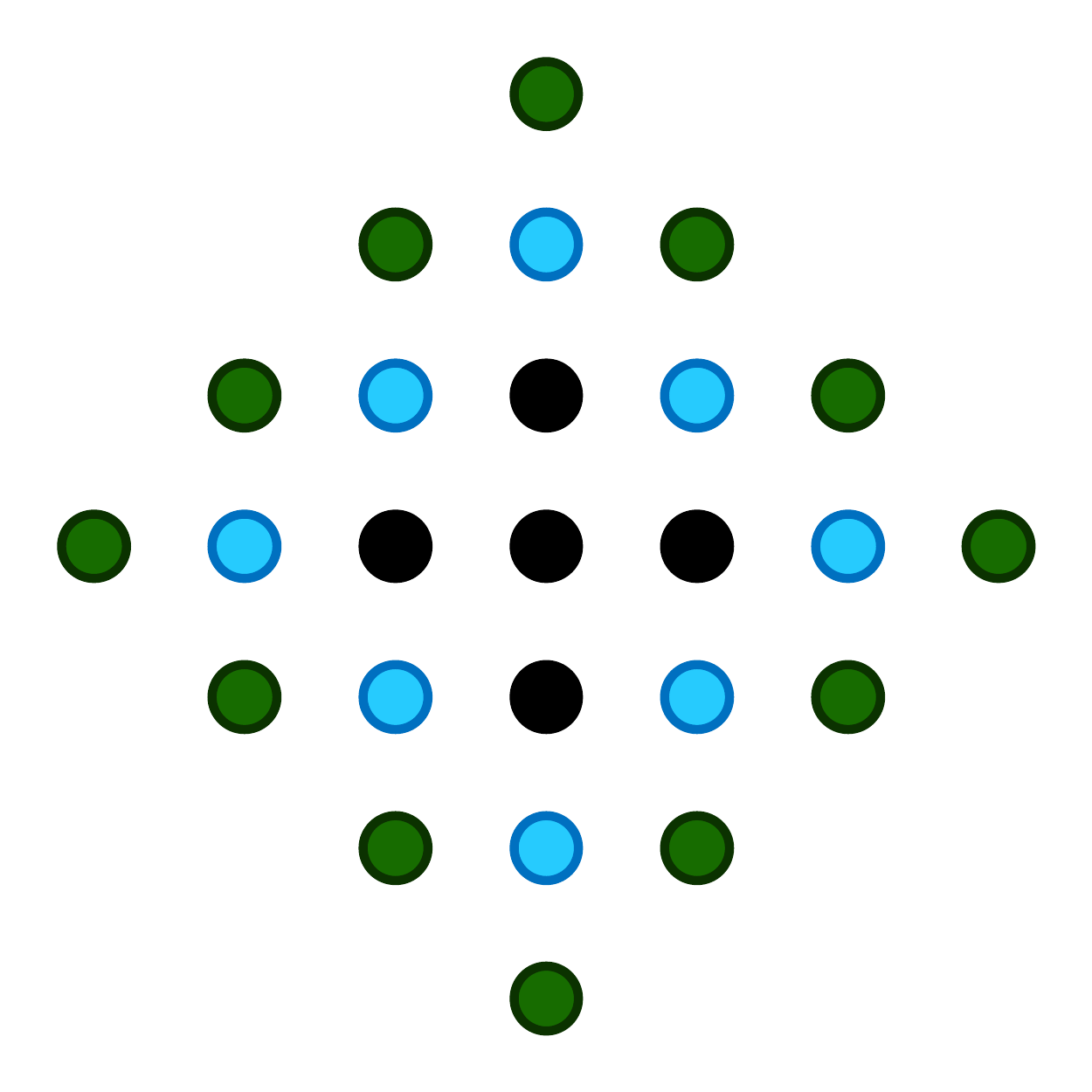}
}
\caption{GHZ state preparation steps. The light blue circles denote the control atoms, the dark green circles denote the target atoms, and the white and black circles indicate atoms not involved in a given step. The light blue and black atoms together are in a GHZ state. After each step, the new GHZ state includes (a) 5, (b) 13, or (c) 25 atoms. \label{GHZfig}}
\end{figure}

The GHZ state preparation protocol is as follows: Initially, all atoms are in a square lattice in the $|0\rangle$ state except for a single atom in the $(|0 \rangle + |1 \rangle)/\sqrt{2}$ state. This single atom will be the control atom while its four nearest neighbors are target atoms. A Hadamard gate is applied to the target qubits, taking them to the $(|0\rangle + |1 \rangle)\sqrt{2}$ state, upon which a C$_1$Z$^3$ gate is applied. A Hadamard gate is applied to the target qubits once more, ending the first step and creating a 5-atom GHZ state. For the subsequent steps, the outermost atoms of the GHZ state are controls while their nearest neighbors outside of the GHZ state are targets. These steps are illustrated in Fig.~\ref{GHZfig}.

The resulting error for each step is
\begin{equation}
\epsilon = (N_c + N_t) \frac{\pi}{4 \Omega \tau} + N_t \langle V_b^{-2} \rangle \Omega^2,
\end{equation}
which has a minimum
\begin{equation}
\epsilon = \frac{3 \pi^{2/3} N_t^{1/3} (N_c+N_t)^{2/3}}{4 (v \tau)^{2/3}},
\end{equation}
where $N_c$ ($N_t$) is the number of control (target) atoms and $\langle V_b^{-2} \rangle = v^{-2}$ is the average value of $V_b^{-2}$ for the target atoms. We have dropped contributions from vdW interactions since they can be made negligible with suitable dressing. For the first, second, and third steps, $\langle (V_{nn}/V_b)^{2} \rangle = 1, .44, .32$, where $V_{nn}$ is the nearest-neighbor interaction. This continues to decrease before reaching a limit of $\langle V_b^{-2} \rangle = .196$ for large steps. 

Based on the dressing of Fig.~3 of the main text, where vdW interactions are made negligible, the maximal nearest-neighbor control-target interaction is $2 \pi \times 2.7$ MHz, which is 80 times smaller than the smallest microwave Rabi frequency. A factor of 10 is to ensure the microwave fields are stronger than the undressed dipole-dipole interactions while the factor of 8 is due to a reduction in the dressed dipole-dipole interactions compared to the undressed dipole-dipole interactions due to the dressing. For $\tau_{c/t} \approx .44$ ms, the errors in the GHZ state preparation are 2\%, 4.5\%, and 7.8\% for 5-, 13-, and 25-atom GHZ states, respectively.